\documentclass[12pt]{article}
\usepackage{epsf}
\usepackage{amsmath}
\usepackage{graphics}
\usepackage{graphicx}

\renewcommand{\thefootnote}{\#\arabic{footnote}}

\newcommand{\gsim}{ \mathop{}_{\textstyle \sim}^{\textstyle >} }
\newcommand{\lsim}{ \mathop{}_{\textstyle \sim}^{\textstyle <} }

\begin{document}

\setcounter{footnote}{0}
\begin{titlepage}

\begin{center}


\vskip .5in

{\Large \bf
Electric Dipole Moments in PseudoDirac Gauginos}

\vskip .45in

{\large
Junji Hisano, Minoru Nagai, Tatsuya Naganawa and Masato Senami
}

\vskip .45in

{\em
Institute for Cosmic Ray Research, \\
University of Tokyo, 
Kashiwa 277-8582, Japan}

\end{center}

\vskip .4in

\begin{abstract}
  The SUSY CP problem is one of serious problems in construction of
  realistic supersymmetric standard models. We consider the problem in
  a framework in which adjoint chiral multiplets are introduced and
  gauginos have Dirac mass terms induced by a U(1) gauge interaction
  in the hidden sector. This is realized in hidden sector models
  without singlet chiral multiplets, which are favored from a recent
  study of the Polonyi problem.  We find that the dominant
  contributions to electron and neutron electric dipole moments (EDMs)
  in the model come from phases in the supersymmetric adjoint mass
  terms.  When the supersymmetric adjoint masses are suppressed by a
  factor of $\sim 100$ compared with the Dirac ones, the electron and
  neutron EDMs are suppressed below the experimental bound even if the
  SUSY particle masses are around 1 TeV. Thus, this model works as a
  framework to solve the SUSY CP problem.  \vspace{1cm}
\end{abstract}

\end{titlepage}

\renewcommand{\thepage}{\arabic{page}}
\setcounter{page}{1}
\renewcommand{\thefootnote}{\#\arabic{footnote}}
\renewcommand{\theequation}{\thesection.\arabic{equation}}

\setcounter{equation}{0}
\section{Introduction}

Origin of gaugino masses is one of the important issues in
supersymmetric (SUSY) extension of the standard model (SM). In the
minimal supergravity model \cite{Nilles:1983ge}, a hidden sector is
introduced, and its interactions with the observable sector are assumed
to be suppressed by the gravitational scale. Majorana gaugino mass
terms in the minimal SUSY SM (MSSM) are derived by a singlet chiral
multiplet with non-vanishing $F$-component in the hidden sector.
However, introduction of the singlet chiral multiplet makes this model
in trouble.

First, other $F$-term SUSY breaking terms in the scalar potential, the
$A$ and $B$ terms, are also generated by the singlet chiral multiplet,
and they have $O(1)$ phases relative to gaugino masses
generically. They contribute to the electron and neutron electric
dipole moments (EDMs) beyond the experimental bounds unless the SUSY
particle masses are heavier than several TeV. This is called as the
SUSY CP problem \cite{susycp,nath}.

Second is a cosmological problem called as the Polonyi problem
\cite{Coughlan:1983ci}.  This problem was considered not to be serious
when the Polonyi field gets a mass of the dynamical SUSY breaking
scale. However, it is recently pointed out that the Polonyi problem is
more serious in hidden sector models in which a singlet chiral
multiplet acquires $F$-term vacuum expectation value (VEV) to generate
the gaugino masses \cite{Ibe:2006am}. The linear term of the singlet
field is allowed in the K$\ddot{\rm a}$hler potential.  When the
Hubble parameter during inflation is not small enough, the linear term
destabilizes the minimum of the singlet scalar (Polonyi) field during
the inflation, and the Polonyi problem is revived.

Therefore, it is one of the ways to construct a realistic SUSY SM to
assume that the hidden sector has no singlet chiral field.  Even if
the hidden sector have no singlet chiral field, Majorana gaugino mass
terms are generated by the pure supergravity (anomaly mediation)
effect \cite{anomaly}. However, they are suppressed by one-loop
factors compared with the gravitino mass. When the particle contents
in the SUSY SM are minimal, the $B$ parameter in the Higgs potential is
of the order of the gravitino mass. Thus, we need to extend this model
furthermore.

An alternative way to generate gaugino mass terms is introduction of
adjoint chiral superfields into the SUSY SM \cite{Hall:1990hq}. When the
hidden sector has a U(1) gauge multiplet whose $D$ component is
non-vanishing, Dirac mass terms with gauginos and the adjoint fermions
are generated \cite{Fox:2002bu,Nelson:2002ca}.  Their sizes can be
comparable to the gravitino mass. Since the gaugino Majorana mass
terms are also generated by the anomaly mediation effect, the
gauginos are pseudoDirac fermions.  We call this model as PseudoDirac
Gaugino (PDG) model.

In this paper we discuss the electron and neutron EDMs in the PDG
model.  The Dirac gaugino mass term is U(1)$_R$ invariant, and it is
pointed out that the EDMs are suppressed in a U(1)$_R$ symmetric limit
\cite{Hall:1990hq}, in which the Majorana gaugino mass terms, the
supersymmetric Higgs mass term, and the $A$ terms are vanishing.  The
U(1)$_R$ breaking terms are introduced by the supergravity effects in
the PDG model. We found that the EDMs are still suppressed when the
supersymmetric adjoint multiplet mass is small.  Thus, the PDG model
is a framework to solve the SUSY CP problem.

This paper is organized as follows. In next section we review the PDG
model in detail, and also discuss the SUSY particle mass spectrum.
In Section 3 we discuss the CP violation and the EDMs derived from it.
Section 4 is devoted to Conclusions and Discussion.

\section{Model}

As mentioned in Introduction, we assume that the hidden sector has no
singlet chiral multiplet, and that the interactions with observable
sector are suppressed by the gravitational scale $M_{\star}$ $(\equiv
2.4\times 10^{18}$ GeV).  Furthermore, it is assumed for simplicity
that the hidden sector does not have fields with VEV$\sim M_\star$.
In this case, the Majorana gaugino masses in the SUSY SM are derived
by the anomaly mediation effect, and they are one-loop suppressed
compared with the gravitino mass ($m_{3/2}$) as
\begin{eqnarray}
\mu_{\tilde{g}_i}&=& \frac{\beta_i(g_i^2)}{2g_i^2}F_\phi.
\label{Mgaugino}
\end{eqnarray}
Here, $\beta_i(g_i^2)$ $(i=Y,2,3)$ are the beta functions for U(1)$_Y$,
SU(2)$_L$ and SU(3)$_C$ gauge coupling constants squared and $F_\phi$
is proportional to $m_{3/2}$.

On the other hand, the scalar masses in the SUSY SM are comparable to
the gravitino mass unless the K$\ddot{\rm a}$hler potential has a
sequestering form between the hidden and observable sectors.  We
assume that the sfermion mass universality is imposed by some flavor
symmetries or underlying physics.

In order to give sizable masses to the gauginos, we introduce 
adjoint chiral multiplets to each gauge multiplet in the SUSY SM. The
adjoint fermions and the gauginos get Dirac mass terms by a VEV of the
$D$ component of a U(1) gauge field in the hidden sector. The kinetic
terms of the adjoint and gauge multiplets and their interaction with
the U(1) gauge field are given as follows,
\begin{eqnarray}
{\cal L}&=&
\int d^2\theta\left\{
\sum_{i=Y,2,3}
\left(
{\rm Tr}
[\frac{1}{4k_ig_i^2} {\cal W}_i^{\alpha} {\cal W}_{i\alpha}]
-\frac{\sqrt{2}c_i}{k_ig_i^2M_\star}{\rm Tr}
[{\cal W}_i^{\alpha} {\cal W}^{\prime}_\alpha \Sigma_i]
\right)+{h.c.}
\right\}
\nonumber\\
&&+\int d^4\theta 
\sum_{i=Y,2,3}
\frac{1}{k_ig_i^2}{\rm Tr}[\Sigma_i^\dagger e^{-2 V_i} \Sigma_i e^{2 V_i}].
\label{dirac}
\end{eqnarray}
Here, ${\cal W}_i^{\alpha}$ and $\Sigma_i$ are gauge field strengths
and adjoint multiplets, respectively, and their interaction with the
U(1) gauge field in the hidden sector, ${\cal W}^{\prime}_\alpha$, is
suppressed by the gravitational scale $M_\star$.  When the U(1) gauge
multiplet in the hidden sector get the $D$-component VEV $\langle
{\cal W}^{\prime}_\alpha \rangle = D\theta_\alpha$, the Dirac masses
for the gauginos and adjoint fermions are given as $M_{D_i}=
c_iD/M_\star$. We assume $D/M_\star\sim m_{3/2}$ so that the gaugino
masses are comparable to the scalar masses in the SUSY SM.

The U(1)$_R$ charges for the adjoint multiplets are zero in this
model, as expected from Eq.~(\ref{dirac}). Thus, their supersymmetric
mass terms vanish in a U(1)$_R$ symmetric limit, similar to those of
the Higgs multiplets in the SUSY SM, $H_1$ and $H_2$.  The
supergravity effect also may generate the mass terms when following
terms are present in the K$\ddot{\rm a}$hler potential
\cite{Giudice:1988yz,Pomarol:1999ie},
\begin{eqnarray}
{\cal L} 
&=&
-\int d^4\theta
\frac{\phi^\dagger}{\phi}
\left\{
c_H H_1H_2
+
\sum_{i=Y,2,3} \frac{c_{\Sigma_i}}{2k_ig_i^2} {\rm Tr}[\Sigma_i^2]
\right\}+{h.c.}.
\label{superhs}
\end{eqnarray}
Here, $\phi(=1+F_\phi \theta^2)$ is the compensator chiral multiplet
in supergravity.  From Eq.~(\ref{superhs}), the supersymmetric Higgs
$(\mu_H)$ and adjoint masses $(\mu_{\Sigma_i})$ and their $B$
parameters ($B_H$ and $B_{\Sigma_i}$) are given as
\begin{eqnarray}
\mu_H&=&c_H F^\star_\phi,\nonumber\\
B_H&=&-F_\phi, \nonumber\\
\mu_{\Sigma_i}&=&c_{\Sigma_i}F^\star_\phi,\nonumber\\
B_{\Sigma_i}&=&-F_\phi. 
\label{bmu}
\end{eqnarray}
When the Higgs and adjoint chiral multiplets have $M_\star$-suppressed
interactions with a non-singlet chiral multiplet $Z$ in the hidden
sector as
\begin{eqnarray}
\int d^4\theta \frac{Z^\dagger Z}{M_\star^2}
\left\{
c^\prime_H H_1H_2
+
\sum_{i=Y,2,3} \frac{c^\prime_{\Sigma_i}}{2k_ig_i^2} {\rm Tr}[\Sigma_i^2]
\right\}+{h.c.},
\label{superhs1}
\end{eqnarray}
their $B$ parameters  and supersymmetric masses are changed from 
Eqs.~(\ref{bmu}).
However, if the VEV for $Z$ is not as large as $\sim M_\star$,
the supergravity contributions in Eq.~(\ref{bmu}) are still dominant
in the supersymmetric masses.

The sizes of $\mu_{\Sigma_i}$ and $\mu_H$ depend on underlying
physics. Even if $c_{\Sigma_i}$ and $c_{H}$ are zero at tree level,
they may be induced radiatively.  For example, when the massive chiral
multiplets $X$ and $\bar{X}$ have an interaction with $\Sigma_i$, $f X\bar{X}
\Sigma_i$, $c_{\Sigma_i}$ is generated at one-loop level as
$c_{\Sigma_i}\sim f^2/(4\pi)^2$, and it is not suppressed by the mass
of $X$ and $\bar{X}$. Thus, $c_{\Sigma_i}$ and $c_{H}$ are expected to
be larger than $O(10^{-(2-3)})$. In the following we assume that
$M_{D_i}\gsim \mu_{\Sigma_i}$ so that the gauginos are pseudoDirac.

The successful gauge coupling unification in the MSSM is spoiled in
the PDG model, since the adjoint chiral multiplets are introduced in
the SUSY SM. In order to recover the unification, it is pointed out in
Ref.~\cite{Fox:2002bu,Nelson:2002ca} that the adjoint chiral
multiplets are embedded in complete multiplets under the GUT gauge
group and the additional fields in the multiplets, which are called as
bachelor fields, are introduced. The candidates for the GUT gauge
group are SU(5) and SU(3)$^3$. The low-energy gauge coupling constants
favor the bachelor masses $(M_B)$ are $10^{(5-9)}$ GeV and the GUT
scale is $10^{(17-18)}$ GeV. It is economical if the GUT scale is
close to the gravitational scale.

The low-energy mass spectrum in the PDG model is affected by the
radiative correction, and it depends on detail of the models between
the weak scale and the gravitational scale. We introduce the bachelor
fields as a working hypothesis and take the GUT scale to be $M_\star$
in the following. The contribution of the adjoint and
bachelor fields to the beta functions for gauge coupling constants
($b_B$) is 5(3) in the SU(5) model (the SU(3)$^3$ model).  
The Dirac gaugino masses at low energy are given from the one-loop
renormalization group (RG) equations as
\begin{eqnarray}
\frac{M_{D_Y}}{g_Y}(m_{SUSY})&=&
\frac{M_{D_Y}}{g_Y}(M_\star), \nonumber\\
\frac{M_{D_2}}{g_2}(m_{SUSY})&=&
\left(\frac{\alpha_2(m_{SUSY})}{\alpha_2(M_{B})}\right)^{-\frac23}
\left(\frac{\alpha_2(M_{B})}{\alpha_2(M_{\star})}\right)^{-\frac2{1+b_B}}
\frac{M_{D_2}}{g_2}(M_\star), \nonumber\\
\frac{M_{D_3}}{g_3}(m_{SUSY})&=&
\left(\frac{M_{B}}{m_{SUSY}}\right)^{\frac{3\alpha_3(m_{SUSY})}{2\pi}}
\left(\frac{\alpha_3(M_{B})}{\alpha_3(M_{\star})}\right)^{-\frac3{-3+b_B}}
\frac{M_{D_3}}{g_3}(M_\star).
\end{eqnarray}
When $b_B= 3$, the SU(3) gaugino mass becomes 
\begin{eqnarray}
\frac{M_{D_3}}{g_3}(m_{SUSY})&=&
\left(\frac{M_{\star}}{m_{SUSY}}\right)^{\frac{3\alpha_3(m_{SUSY})}{2\pi}}
\frac{M_{D_3}}{g_3}(M_\star).
\end{eqnarray}
The gaugino mass differences among the gauge groups are larger than
those in the MSSM under the minimal supergravity assumption (cMSSM),
in which the gaugino mass ratios are given by those of squares of the
gauge coupling constants. If $\sqrt{3/5}{M_{D_Y}}/{g_Y}
={M_{D_2}}/{g_2}={M_{D_3}}/{g_3}$ at $M_\star$, the gaugino mass
ratios in the SU(3)$^3$ model are $M_{D_2}/M_{D_Y}\simeq (2.3-2.6)$
and $M_{D_3}/M_{D_Y}\simeq 11$, depending on $M_B$. In the SU(5)
model, the ratios are larger than those in the SU(3)$^3$ model,
$M_{D_2}/M_{D_Y}\simeq (2.3-4.3)$ and $M_{D_3}/M_{D_Y}\simeq (11-31)$.
This is because the gauge coupling constants at $M_\star$ are quite
large when $M_B \sim 10^{(7-8)}$ GeV.

The adjoint supersymmetric masses $\mu_{\Sigma_i}$ also receive
sizable radiative corrections. The following combination including
$\mu_{\Sigma_i}$ is RG invariant at one-loop level,
\begin{eqnarray}
\frac{\mu_{\tilde{g}_i}\mu_{\Sigma_i}}{M_{D_i}^2},
\end{eqnarray}
for $i=Y,2,3$. 

The light Higgs boson is predicted to be lighter in the PDG model,
relatively to the MSSM. It is, unfortunately, contrary to null results
of the Higgs boson searches.  The first reason comes from the quartic
coupling of the Higgs boson given from Eq.~(\ref{dirac}) as
\begin{eqnarray}
\frac18\left(
\frac{m_{A_2}^2}{m_{A_2}^2+4 M_{D_2}^2}g_2^2
+
\frac{m_{A_Y}^2}{m_{A_Y}^2+4 M_{D_Y}^2}g_Y^2\right).
\end{eqnarray}
Here, we take simply $\mu_{\Sigma_i}B_{\Sigma_i}\simeq 0$ and
$\mu_{\Sigma_i}\simeq 0$, and $m_{A_i}(i=Y,2)$ are the SUSY breaking
masses of the adjoint scalars for U(1)$_Y$ and SU(2)$_L$
\cite{Fox:2002bu}. The MSSM quartic coupling is recovered in a case of
$m_{A_i}\gg M_{D_i}$. The quartic coupling is suppressed when
$m_{A_i}\sim M_{D_i}$, and the Higgs boson mass becomes lighter.  The
second reason is that while the large $A$ parameter for top squarks
enhances the radiative correction to the light Higgs boson mass
\cite{Kitano:2006gv}, it is one-loop suppressed in the PDG model.
Since the hidden sector have no singlet fields, the $A$ parameters
come from only the anomaly mediation similar to the Majorana gaugino
masses.  The $A$ parameters for the supersymmetric Yukawa coupling
$\lambda_{flm}$ are given as
\begin{eqnarray}
A_{flm}&=& -(\gamma_f+\gamma_l+\gamma_m)F_\phi
\label{MApara}
\end{eqnarray}
where $\gamma_f$ is the anomalous dimension for a field $f$.
These implies from null results of the light Higgs boson searches that
the large top squark masses ($\gsim$  1 TeV) are required so that the
radiative correction to the Higgs boson mass is enhanced.  

In the PDG model the radiative corrections to the scalar masses due to
non-vanishing Dirac gaugino masses are finite and are not enhanced by
a logarithm. This nature is called as ``supersoft'' \cite{Fox:2002bu}.
The first and second-generation sfermions do not receive large
radiative corrections to the mass terms when the gaugino masses are
comparable to the sfermion masses and the U(1)$_Y$ $D$-term
contribution to the RG equations is zero. The third generation
sfermions and Higgs boson masses get RG effects by the large Yukawa
interaction.

\section{Electron and neutron EDMs}

First, we review electron and neutron EDMs in the MSSM briefly,
and show the constraints on the CP phases in the parameters in the
model.  In the MSSM the new sources of CP violation in the
flavor-conserving terms are phases in the gaugino masses, $A$ and $B$
parameters, and the supersymmetric Higgs mass. The physical CP phases
are following two due to degrees of freedom for rephasing, U(1)$_R$
and U(1)$_{PQ}$,
\begin{eqnarray}
\phi_{\mu_{\tilde{g}}}\equiv {\rm arg}(\mu_{H}\mu_{\tilde{g}_i}(\mu_HB_H)^\star),~~
\phi_{A_{f}}\equiv {\rm arg}(\mu_{\tilde{g}_i} A_f^\star).
\label{MSSMcp}
\end{eqnarray}
Here the gaugino mass unification in GUTs and universality of
$A$ parameters are assumed.

The relevant CP-violating operators for the electron and neutron EDMs are
following up to dimension six \cite{Pospelov:2005pr},
\begin{eqnarray}
{\cal L}_{\rm (C)EDM}&=&
-\sum_{f=e,u,d} i \frac{d_f}{2} 
\bar{f} (F \cdot \sigma) \gamma_5 f
\nonumber\\
&&+\overline{\theta} \frac{\alpha_s}{8\pi} G\tilde{G}
-\sum_{q=u,d} i\frac{d^c_q}2 \bar{q}(g_s G\cdot \sigma) \gamma_5 q
+\frac13 \omega G\tilde{G} G,
\label{eff_cp}
\end{eqnarray}
where $F_{\mu\nu}$ and $G_{\mu\nu}$ are the electromagnetic and the
SU(3)$_C$ gauge field strengths, respectively. The terms in the second
line are the flavor-conserving CP-violating terms in QCD, and the
first term is the QCD theta term. Here, we simply assume Peccei-Quinn
symmetry for it. The second and third terms are quark chromoelectric
dipole moments (CEDMs) and  the Weinberg operator, respectively.

The current experimental bound on the electron EDM is $|d_e|<
1.6\times 10^{-27}~e{\rm cm}$ from the $^{205}$Tl EDM measurement
\cite{Regan:2002ta}. The neutron EDM bound is recently improved as
$|d_n|<2.9\times 10^{-26}~e{\rm cm}$ \cite{Baker:2006ts}.  The
evaluation of the neutron EDM from the parton level is still a
difficult task. Here, we use a formula derived from the QCD sum rule
\cite{sumrule}, in which the valence quark contribution to the EDM is
evaluated\footnote {
  The sea quarks, including strange quark, may contribute to the
  neutron EDM sizable \cite{Zhitnitsky:1996ng}. In this paper we
  ignore them for simplicity since the evaluation still has
  uncertainties.  The mercury EDM \cite{Romalis:2000mg} is also
  sensitive to the quark CEDMs, and the bound is as stringent as the
  neutron one though it is also expected to suffer from larger
  hadronic uncertainties than neutron one due to the nuclear dynamics
  \cite{Pospelov:2005pr}.
},
\begin{eqnarray}
\frac{d_n}e= (1\pm 0.5)\frac{|\langle \bar{q}q \rangle |}{(225~{\rm MeV})^3}
\left(
1.1\left(d^c_d+0.5d_u^c\right)+1.4\left(\frac{d_d}e-0.25\frac{d_u}e\right)
\right).
\label{sumrulenedm}
\end{eqnarray}
The Weinberg operator contribution to the neutron EDM is also evaluated as 
\cite{Demir:2002gg}
\begin{eqnarray}
\frac{d_n}e \simeq \omega \times 22~{\rm MeV},
\end{eqnarray}
though it has more uncertainties.

In the MSSM the electron EDM is given as \begin{eqnarray}
\frac{d_e}e
&=&
\sum_{i=Y,2}
{\rm Im}\left[\mu_{\tilde{g}_i} \mu_H \right]
d_i^{(e)}(|\mu_{\tilde{g}_i}|^2)
+{\rm Im}\left[\mu_{\tilde{g}_Y}A_l^\star \right]
d_Y^{(e)\prime}(|\mu_{\tilde{g}_Y}|^2),
\label{eedm}
\end{eqnarray}
while the quark EDMs and CEDMs are given as 
\begin{eqnarray}
\frac{d_q}e
&=&
\sum_{i=Y,2,3}
{\rm Im}\left[\mu_{\tilde{g}_i} \mu_H \right]
d_i^{(q)}(|\mu_{\tilde{g}_i}|^2)
+
\sum_{i=Y,3}
{\rm Im}\left[\mu_{\tilde{g}_i} A_q^\star \right]
d_i^{(q)\prime}(|\mu_{\tilde{g}_i}|^2),
\nonumber\\
d_q^c
&=&
\sum_{i=Y,2,3}
{\rm Im}\left[\mu_{\tilde{g}_i} \mu_H \right]
\bar{d}_i^{(q)}(|\mu_{\tilde{g}_i}|^2)
+
\sum_{i=Y,3}
{\rm Im}\left[\mu_{\tilde{g}_i} A_q^\star \right]
\bar{d}_i^{(q)\prime}(|\mu_{\tilde{g}_i}|^2),
\label{qedm}
\end{eqnarray}
where $q=u,d$ and we take a basis in which $\mu_H B_H$ is real.  The
mass functions for EDMs, $d^{(f)}_i(x)$ and $d^{(f)\prime}_i(x)$
$(f=e,u,d)$, and those for CEDMs, $\bar{d}^{(f)}_i(x)$ and
$\bar{d}^{(f)\prime}_i(x)$ $(f=u,d)$, are defined for convenience to
evaluate the EDMs and CEDMs in the PDG model later. The explicit forms
in a limit $m_{SUSY}\gg m_Z$ are given in Appendix. More complete
formulae can be derived from those in Ref.~\cite{nath}. The Weinberg
operator contribution to the neutron EDM is subdominant in the MSSM
except for cases in some specific mass spectrum, and then it is ignored here.

In a limit of common SUSY breaking parameters $m_{SUSY}$, the EDMs and CEDMs
are
\begin{eqnarray}
\frac{d_e}e
&=&
-\left(\frac{5\alpha_2}{24}
+\frac{\alpha_Y}{24}
\right)
\frac{m_e\tan\beta}{4\pi m_{SUSY}^2}
\sin\phi_{\mu_{\tilde{g}}}
-\frac{\alpha_Y}{12}
\frac{m_e}{4\pi m_{SUSY}^2}
\sin\phi_{A_{l}},
\label{mssmde}
\end{eqnarray}
and
\begin{eqnarray}
\frac{d_d}e
&=&
-\left(
\frac{2\alpha_3}{27}
+\frac{7\alpha_2}{24}
-\frac{11\alpha_Y}{648}
\right)
\frac{m_d\tan\beta }{4\pi m_{SUSY}^2}
\sin\phi_{\mu_{\tilde{g}}}
-
\left(\frac{2\alpha_3}{27}
-\frac{\alpha_Y}{324}
\right)
\frac{m_d}{{4\pi}
m_{SUSY}^2}
\sin\phi_{A_d},
\nonumber\\
d_d^c
&=&
-
\left(
\frac{5\alpha_3}{18}
+\frac{\alpha_2}{8}
+\frac{11\alpha_Y}{216}
\right)
\frac{m_d\tan\beta }{4 \pi m_{SUSY}^2}
\sin \phi_{\mu_{\tilde{g}}}
-
\left(
\frac{5\alpha_3}{18}
+\frac{\alpha_Y}{108}
\right)
\frac{m_d}{{4\pi}m_{SUSY}^2}
\sin \phi_{A_d},
\nonumber\\
\frac{d_u}e
&=&
+
\left(
\frac{4\alpha_3}{27}
+\frac{\alpha_2}{4}
-\frac{5\alpha_Y}{324}
\right)
\frac{m_u\tan^{-1}\beta }{{4\pi}m_{SUSY}^2}
\sin \phi_{\mu_{\tilde{g}}}
+
\left(
\frac{4\alpha_3}{27}
+\frac{\alpha_Y}{81}
\right)
\frac{m_u}{4\pi m_{SUSY}^2}
\sin \phi_{A_u},
\nonumber\\
d_u^c
&=&
-
\left(
\frac{5\alpha_3}{18}
+\frac{\alpha_2}{8}
+\frac{5\alpha_Y}{216}
\right)
\frac{m_u\tan^{-1}\beta }{4\pi m_{SUSY}^2}
\sin \phi_{\mu_{\tilde{g}}}
-
\left(
\frac{5\alpha_3}{18}
-\frac{\alpha_Y}{54}
\right)
\frac{m_u}{{4\pi}
m_{SUSY}^2}
\sin \phi_{A_u}.
\nonumber\\
\label{mssmdq}
\end{eqnarray}
From these equations, the phases are constrained as
\begin{eqnarray}
|\sin\phi_{\mu_{\tilde{g}}}| <
\left(\frac{m_{SUSY}}{6.1~{\rm TeV}}\right)^{2}
\left({\frac{\tan\beta}{10}}\right)^{-1},~~
|\sin\phi_{A_f}| <
\left(\frac{m_{SUSY}}{0.65~{\rm TeV}}\right)^{2},
\end{eqnarray}
from the experimental bound on the electron EDM, and 
\begin{eqnarray}
|\sin\phi_{\mu_{\tilde{g}}}| <
\left(\frac{m_{SUSY}}{12~{\rm TeV}}\right)^{2}
\left({\frac{\tan\beta}{10}}\right)^{-1},~~
|\sin\phi_{A_f}| <
\left(\frac{m_{SUSY}}{3.7~{\rm TeV}}\right)^{2},
\label{nedm}
\end{eqnarray}
from neutron one.  Here, we use $\alpha_3(m_Z)=0.1176$,
$\alpha^{-1}(m_Z)=127.918$, $\sin^2\theta_W(m_Z)=0.23122$ \cite{PDG},
$m_d(m_Z)=4.1 $ MeV and $m_u(m_Z)=2.4$ MeV\footnote{
  These values are evaluated from  $m_d(1~{\rm GeV})=8.9$ MeV and $m_u(1~{\rm
    GeV})=5.1$ MeV \cite{Gasser:1984gg}.
} and the QCD corrections to the dipole operators are ignored.  The
center value in Eq.~(\ref{sumrulenedm}) is taken in evaluation of
Eq.~(\ref{nedm}). The down quark contribution tends to dominate in the
neutron EDM. Especially, it is significant when
$\sin\phi_{\mu_{\tilde{g}}}\ne 0$. It is found from these bounds that
the SUSY particles should be larger than several TeV when the CP
phase is $O(1)$.

In the PDG model the complex parameters introduced are
\begin{eqnarray}
M_{D_i},~\mu_{\tilde{g}_i},~\mu_{\Sigma_i},~B_{\Sigma_i} ~~(i=Y,2,3),\nonumber\\
\mu_H,~ B_H,~~A_f~~(f=u,d,l).
\end{eqnarray}
Degrees of freedom for rephasing come from phase transformations of
adjoint fields in addition to U(1)$_R$ and U(1)$_{PQ}$ in the
MSSM. Thus, the physical CP phases are four, including
$\phi_{\mu_{\tilde{g}}}$ and $\phi_{A_{f}}$ in Eq.~(\ref{MSSMcp}),
when assuming GUT and universality of the $A$ terms. The new phases in
the PDG model are
\begin{eqnarray}
\phi_{\mu_{\Sigma}}\equiv {\rm arg}(\mu_H\mu^\star_{\Sigma_i} (\mu_HB_H)^\star
M_{D_i}^{2}),~~
\phi_{B_{\Sigma}}\equiv {\rm arg}(\mu_{\Sigma_i} B_{\Sigma_i} M_{D_i}^{\star2}).
\end{eqnarray}
In the following we take a basis in which $M_{D_i}$ and
$\mu_H B_H$ are real for convenience.

When the hidden sector does not have singlet chiral multiplets, the
$A$ parameters and the Majorana gaugino masses are dominated by the
anomaly mediation contribution as in Eqs.~(\ref{Mgaugino}) and
(\ref{MApara}), and their phases are automatically aligned. That is,
$\phi_{A_{f}}=0$. On the other hand, the $B$ parameters still have
other contributions even in the case, and $\phi_{B_{\Sigma}}$ and
$\phi_{\mu_{\tilde{g}}}$ do not necessarily vanish.  The adjoint
scalar bosons are not directly coupled to quarks and leptons. The CP
phase $\phi_{B_{\Sigma}}$ contributes to the EDMs at three-loop level
via the Weinberg operator \cite{Hall:1990hq}.  We evaluate the EDMs in
$\phi_{\mu_{\tilde{g}}}\ne 0$ and $\phi_{\mu_{\Sigma}}\ne 0$,
first. The Weinberg operator contribution in $\phi_{B_{\Sigma}}\ne 0$
is discussed later.

The adjoint fermions are also coupled to quarks and leptons via the
gauginos. Thus, when $\mu_{\tilde{g}_i}$ and $\mu_{{\Sigma}_i}
\ll M_{D_i}$, the one-loop contributions to the EDMs and CEDMs are
derived from those in the MSSM as
\begin{eqnarray}
\frac{d_e}e
&=&
\sum_{i=Y,2}\left\{
{\rm Im}\left[\mu_{\tilde{g}_i} \mu_H \right]
\left(1+M_{D_i}^2\frac{\partial}{\partial M_{D_i}^2}\right)
d_i^{(e)}(M_{D_i}^2)
+
{\rm Im}\left[\mu_{\Sigma_i}^\star \mu_H \right]
M_{D_i}^2\frac{\partial}{\partial M_{D_i}^2}
d_i^{(e)}(M_{D_i}^2)
\right\}
\nonumber\\
&+&
{\rm Im}\left[\mu_{\tilde{g}_Y} A_l^\star \right]
\left(1+M_{D_Y}^2\frac{\partial}{\partial M_{D_Y}^2}\right)
d_Y^{(e)\prime}(M_{D_Y}^2)
-
{\rm Im}\left[\mu_{\Sigma_Y} A_l \right]
M_{D_Y}^2\frac{\partial}{\partial M_{D_Y}^2}
d_Y^{(e)\prime}(M_{D_Y}^2), 
\nonumber\\
\frac{d_q}e
&=&
\sum_{i=Y,2,3}\left\{
{\rm Im}\left[\mu_{\tilde{g}_i} \mu_H \right]
\left(1+M_{D_i}^2\frac{\partial}{\partial M_{D_i}^2}\right)
d_i^{(q)}(M_{D_i}^2)
+
{\rm Im}\left[\mu_{\Sigma_i}^\star \mu_H \right]
M_{D_i}^2\frac{\partial}{\partial M_{D_i}^2}
d_i^{(q)}(M_{D_i}^2)
\right\}
\nonumber\\
&+&
\sum_{i=Y,3}\left\{
{\rm Im}\left[\mu_{\tilde{g}_i} A_q^\star \right]
\left(1+M_{D_i}^2\frac{\partial}{\partial M_{D_i}^2}\right)
d_i^{(q)\prime}(M_{D_i}^2)
-
{\rm Im}\left[\mu_{\Sigma_i} A_q \right]
M_{D_i}^2\frac{\partial}{\partial M_{D_i}^2}
d_i^{(q)\prime}(M_{D_i}^2)
\right\},
\nonumber\\
d_q^c
&=&
\sum_{i=Y,2,3}\left\{
{\rm Im}\left[\mu_{\tilde{g}_i} \mu_H \right]
\left(1+M_{D_i}^2\frac{\partial}{\partial M_{D_i}^2}\right)
\bar{d}_i^{(q)}(M_{D_i}^2)
+
{\rm Im}\left[\mu_{\Sigma_i}^\star \mu_H \right]
M_{D_i}^2\frac{\partial}{\partial M_{D_i}^2}
\bar{d}_i^{(q)}(M_{D_i}^2)
\right\}
\nonumber\\
&+&
\sum_{i=Y,3}\left\{
{\rm Im}\left[\mu_{\tilde{g}_i} A_q^\star \right]
\left(1+M_{D_i}^2\frac{\partial}{\partial M_{D_i}^2}\right)
\bar{d}_i^{(q)\prime}(M_{D_i}^2)
-
{\rm Im}\left[\mu_{\Sigma_i} A_q \right]
M_{D_i}^2\frac{\partial}{\partial M_{D_i}^2}
\bar{d}_i^{(q)\prime}(M_{D_i}^2)
\right\},
\nonumber\\
\end{eqnarray}
($q=u,d)$.  We include the contribution in $\phi_{A_f}\ne 0$ for
completeness though it is subdominant.

The one-loop contributions to the EDMs and CEDMs are suppressed by
$\mu_{\tilde{g}_i}$ or $\mu_{\Sigma_i}$. When taking a common value
for SUSY breaking parameters except for $\mu_{\tilde{g}_i}$ and
$\mu_{\Sigma_i}$, the electron EDM and down quark EDM and CEDM are
given as follows,
\begin{eqnarray}
\frac{d_e}{e}
&=&
-\frac{\alpha_2\delta_{\tilde{g}_2}}{30}
\frac{m_e\tan\beta}{{4\pi}m_{SUSY}^2}
\sin\phi_{\mu_{\tilde{g}}}
+
\left(
\frac{7{\alpha_2\delta_{\Sigma_2}}}{40}
+\frac{{\alpha_Y\delta_{\Sigma_Y}}}{24}
\right)
\frac{m_e\tan\beta}{4\pi m_{SUSY}^2}\
\sin \phi_{\mu_{\Sigma}}, 
\nonumber\\
\frac{d_d}e
&=&
-\left(
\frac{2\alpha_3\delta_{\tilde{g}_3}}{135}
+\frac{\alpha_2\delta_{\tilde{g}_2}}{15}
-\frac{\alpha_Y\delta_{\tilde{g}_Y}}{162}
\right)
\frac{m_d\tan\beta}{4\pi m_{SUSY}^2}
\sin\phi_{\mu_{\tilde{g}}}
\nonumber\\
&& 
+
\left(
\frac{8{\alpha_3\delta_{\Sigma_3}}}{135}
+\frac{9{\alpha_2 \delta_{\Sigma_2}}}{40}
-\frac{7{\alpha_Y \delta_{\Sigma_Y}}}{648}
\right)
\frac{m_d\tan\beta}{{4\pi}m_{SUSY}^2}
\sin\phi_{\mu_{\Sigma}},
\nonumber\\
d_d^c
&=&
+\left(
\frac{2{\alpha_3\delta_{\tilde{g}_3}}}{45}
-\frac{\alpha_2\delta_{\tilde{g}_2}}{20}
-\frac{\alpha_Y\delta_{\tilde{g}_Y}}{54}
\right)
\frac{m_d\tan\beta}{4\pi m_{SUSY}^2}
\sin\phi_{\mu_{\tilde{g}}}
\nonumber\\
&& 
+
\left(
\frac{29 {\alpha_3\delta_{\Sigma_3}}}{90}
+\frac{3{\alpha_2 \delta_{\Sigma_2}}}{40}
+\frac{7{\alpha_Y \delta_{\Sigma_Y}}}{216}
\right)
\frac{m_d\tan\beta}{4\pi m_{SUSY}^2}
\sin\phi_{\mu_{\Sigma}},
\end{eqnarray}
where $\delta_{\tilde{g}_i}=|\mu_{\tilde{g}_i}/m_{SUSY}|$ and
$\delta_{\Sigma_i}=|\mu_{\Sigma_i}/m_{SUSY}|$. We keep terms
proportional to $\tan\beta$ in the equations. Compared with
Eqs.~(\ref{mssmde}) and (\ref{mssmdq}), it is found that
$\delta_{\tilde{g}_i}$ and $\delta_{\Sigma_i}$ should be smaller than
$\sim 0.01$ so that the electron and neutron EDMs are suppressed below
the experimental bounds.

\begin{figure}[t]
\begin{tabular}{cc}
\includegraphics[scale=0.8]{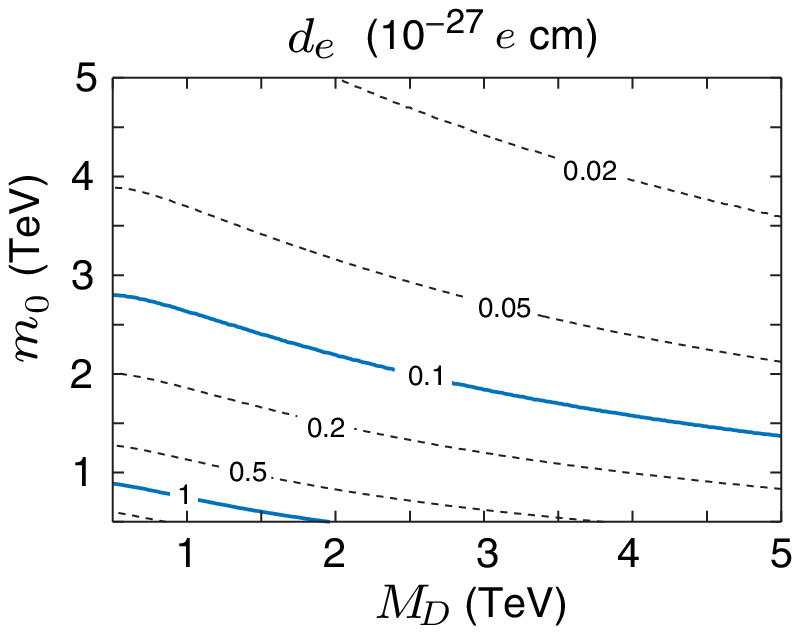}&
\includegraphics[scale=0.8]{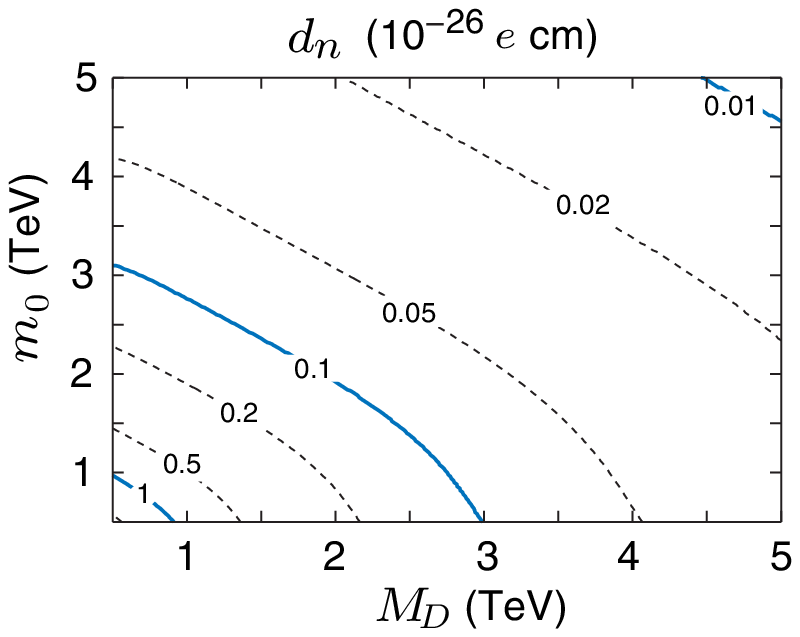} 
\end{tabular}
\caption{\label{fig1} Electron and neutron EDMs in the PDG model as
  functions of Dirac gaugino mass ($M_D$) and sfermion mass ($m_0$) at the
  gravitational scale $M_\star$. We assume the SU(5) model for the
  bachelor fields and the sfermion mass universality. Here,
  $\sin\phi_{\mu_{{\Sigma}}}=1$, $\sin\phi_{\mu_{{\tilde{g}}}}=0$ and
  $|\mu_{\Sigma}/M_{D}|=10^{-3}$ at $M_\star$, $M_B=10^9$~GeV, 
  $\mu_{H}=200~{\rm GeV}$, and $\tan\beta=10$.  }
\end{figure}

\begin{figure}[t]
\begin{tabular}{cc}
\includegraphics[scale=0.8]{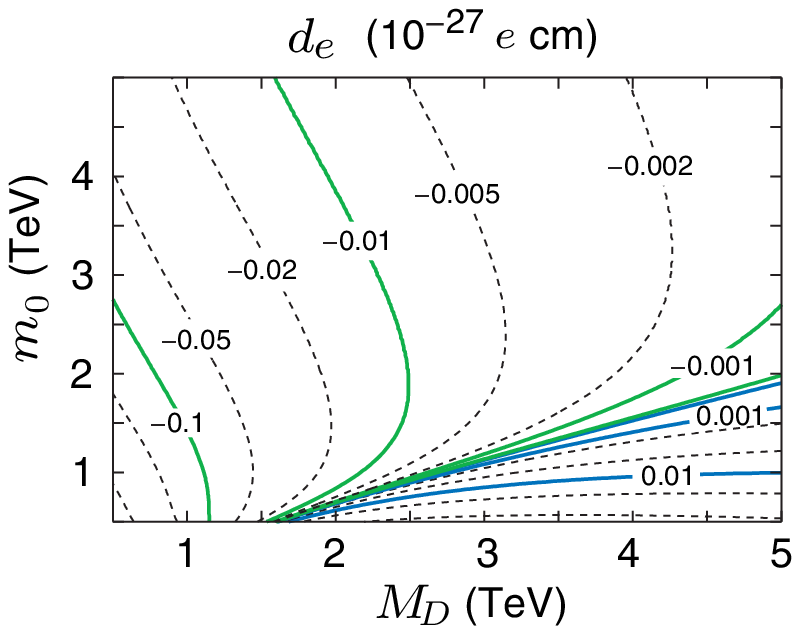}&
\includegraphics[scale=0.8]{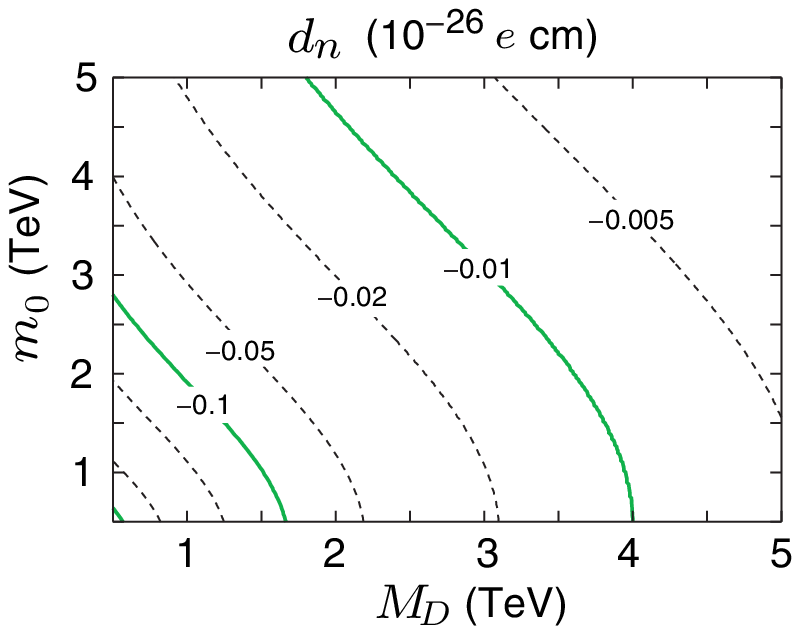} 
\end{tabular}
\caption{\label{fig2} Electron and neutron EDMs in the PDG model as 
  functions of Dirac gaugino mass and sfermion mass at the
  gravitational scale. Here, $F_\phi/M_D=1$, $\mu_{\Sigma_i}=0$ and
  $\sin\phi_{\mu_{\tilde{g}}}=1$.  The other assumptions and input
  parameters are the same as in Fig.~\ref{fig1}.  }
\end{figure}

In Fig.~1 we show the electron and neutron EDMs in the PDG model as
functions of the Dirac gaugino mass ($M_{D}$) and the sfermion mass
($m_0$) at the gravitational scale $M_\star$.  We assume the SU(5)
model for the bachelor fields and the sfermion mass
universality. Here, we evaluate the EDMs in a case that
$|{\mu_{{\Sigma}_i}}/M_D|=10^{-3}$, $\sin\phi_{\mu_{{\Sigma}}}=1$, and
$\sin\phi_{\mu_{{\tilde{g}}}}=0$ at $M_\star$.  In this case,
$|{\mu_{{\Sigma}_3}}/M_D|=0.5\times 10^{-2}$ and
$|{\mu_{{\Sigma}_3}}/M_D|=5.7\times 10^{-2}$ at $m_{SUSY}$ due to the
radiative corrections, and the phases in the adjoint fermion masses,
$\phi_{\mu_{{\Sigma}}}$, are the dominant sources of CP violation in
the EDMs.  Other parameters are
taken as $M_B=10^9$~GeV, $\mu_{H}=200~{\rm GeV}$, and
$\tan\beta=10$. The EDMs are proportional to $\mu_{\Sigma_i}$ in
$\mu_{\Sigma_i}/M_{D}\ll 1$.  It is found from the figure that if
$|{\mu_{{\Sigma}_i}}/M_D|\lsim 0.01$, the EDMs are suppressed
below the experimental bounds even when $m_0$ and $M_{D}$ are smaller
than 1 TeV.

In Fig.~2 the electron and neutron EDMs are shown in a case of
$\mu_{\Sigma_i}=0$, $|F_\phi/M_D|=1$ and
$\sin\phi_{\mu_{{\tilde{g}}}}=1$.\footnote{
  Notice that in the PDG model the one-loop beta function for the
  SU(3)$_C$ gauge coupling constant is zero. Thus, non-vanishing
  $\mu_{\tilde{g}_3}$ is derived at two-loop level, and the gluino
  contribution to the quark (C)EDM in $\sin\phi_{\mu_{{\Sigma}}}= 0$
  is subdominant.
} Other parameters are the same as in Fig.~1. In this case, the phases
in the Majorana gaugino mass terms are the dominant sources of CP
violation in the EDMs. Even if $\mu_{\Sigma_i}$ is suppressed, the
Majorana gaugino masses are derived by the anomaly mediation
effect. Thus, the evaluated EDMs are the lowerbounds on the EDMs in
the PDG model unless accidental cancellation suppress the EDMs. 

The EDMs in the case of $\mu_{\Sigma_i}=0$ and
$\sin\phi_{\mu_{{\tilde{g}}}}=1$ are much suppressed below the current
experimental bounds.  Especially, the electron EDM is suppressed by
the cancellation when $m_0\lsim 2$ TeV and $M_D\gsim 1.5$
TeV. However, the finite values are still predicted so that the future
experiments may cover them.

The neutron EDM measurements will be improved at SNS and ILL in near
future, and it is argued that their sensitivities may reach to $\sim
10^{-28}~e{\rm cm}$ \cite{lepto06}. The deuteron EDM is also planned to
be measured in the storage ring with sensitivity $10^{-29}~e{\rm cm}$,
which corresponds to $d_n\sim 10^{-(29-30)}~e{\rm cm}$
\cite{Semertzidis:2003iq,lepto06}. The electron EDM measurements using
polarizable paramagnetic molecules are aiming to the
sensitivities of $10^{-29}~e{\rm cm}$ \cite{Hudson:2002az,lepto06}.

Finally, we discuss the Weinberg operator induced by non-zero
$\phi_{B_{\Sigma}}$. It is derived at two-loop diagrams, and
roughly evaluated by dimensional counting as
\begin{eqnarray}
\omega~\sim \frac{\alpha_3^2}{(4\pi)^2}
\frac{{\rm Im}[\mu_{\Sigma_3} B_{\Sigma_3}]}{m_{SUSY}^4}.
\end{eqnarray}
Thus, the neutron EDM is derived from it as
\begin{eqnarray}
{d_n}\sim 4\times 10^{-26}~e{\rm cm}\times
\left(\frac{{{\rm Im}[\mu_{\Sigma_3} B_{\Sigma_3}]}}{m_{SUSY}^2}\right)
\left(\frac{m_{SUSY}}{1~{\rm TeV}}\right)^{-2},
\end{eqnarray}
though it also suffers from large hadronic uncertainties.
The Weinberg operator contribution is marginal to the current bound on
the neutron EDM if $\mu_{\Sigma_3} B_{\Sigma_3}$ is not suppressed. 
When the supersymmetric adjoint mass terms and the SUSY breaking term
associated with them come from independent terms, such as in
Eqs.~(\ref{superhs}) and (\ref{superhs1}),
$\mu_{\Sigma_3}B_{\Sigma_3}$ may be comparable to $m_{3/2}^2$ while
$\mu_{\Sigma_3}\ll m_{3/2}$.  In such models, the Weinberg operator
contribution would be dominant in the neutron EDM.

\section{Conclusions and Discussion}

The SUSY CP problem is one of the serious problems in construction of
realistic SUSY SMs. We consider the problem in a framework in which
the gaugino mass terms are Dirac-type ones induced by a U(1) gauge
interaction in the hidden sector. The adjoint chiral multiplets are
introduced to have the Dirac mass terms.  This model is called as the
PDG model in this paper. The setup is also preferable from a viewpoint
of the Polonyi problem, since we do not need to introduce singlet
chiral superfields in the hidden sector in order to derive the gaugino
masses as in the minimal supergravity model.

We find that the dominant contributions to the electron and neutron
EDMs in the PDG model come from the phases in the supersymmetric
adjoint mass terms.  When the supersymmetric adjoint masses are
suppressed by a factor of $\sim 100$ compared with the Dirac ones, the
neutron and electron EDMs are suppressed below the experimental bounds
even if the SUSY particle masses are around 1 TeV. Thus, this model
works as a framework to solve the SUSY CP problem.

The electron and neutron EDMs are suppressed in the PDG model,
however, the predicted values may be covered in the near future
experiments. Even if the supersymmetric adjoint masses are much
suppressed, the Majorana gaugino mass terms, which are induced by the
anomaly mediation effect, may induce large enough neutron EDM to be
observed in near future experiments. In addition to it, the phase in
the $B$ term of the adjoint scalar boson contributes to the Weinberg
operator. It is not necessarily suppressed even if the supersymmetric
adjoint mass is. In the case, the Weinberg operator would be dominant
in the neutron EDM.

In this paper, we consider only the CP-violating terms which are
flavor-conserving. The flavor-violating ones also contribute to the
EDMs in the SUSY SM. Especially, when both the left- and right-handed
sfermions have the flavor-mixing mass terms, the phases in them
contribute to the EDMs at one-loop level. Even if the sfermion
mixing angles are suppressed similar to the CKM ones, the EDMs can
reach to the experimental bounds in the MSSM \cite{flavorcp}.  These
contributions are also suppressed in the PDG model, since the
contributions are proportional to the Majorana bino or gluino masses.

\section*{Acknowledgments}
The works of JH and MS are supported in part by the Grant-in-Aid
for Science Research, Ministry of Education, Science and Culture,
Japan (No.~1554055 and No.~18034002 for JH and No.~18840011 for MS).
Also, that of MN  is supported in part by JSPS.

\appendix

\section{Appendix}

Here, we present mass functions for the electron and quark EDMs and
quark CEDMs in the MSSM. Here, we assume
$|\mu_{\tilde{g}_{2/Y}}\pm\mu_H| \gg m_Z$. Furthermore, we take a
common value $m$ for the sfermion masses.

The mass functions for the electron EDM in Eq.~(\ref{eedm}) are given
as follows,
\begin{eqnarray}
d_2^{(e)}(|\mu_{\tilde{g}_2}|^2)
&=&
\frac{\alpha_2}{4\pi}
m_e \tan\beta
\times
\hat{D}^{(|\mu_{\tilde{g}_2}|^2, |\mu_H|^2)}_{M^2}
\left[
-\frac{1}{4m^2} f_0(M^2/m^2)
-\frac{1}{2m^2} f_1(M^2/m^2)
\right],
\nonumber\\
d_Y^{(e)}(|\mu_{\tilde{g}_Y}|^2)
&=&
\frac{\alpha_Y}{4\pi}
m_e \tan\beta
\times
\hat{D}^{(|\mu_{\tilde{g}_Y}|^2, |\mu_H|^2)}_{M^2}
\left[
-\frac{1}{4 m^2} f_0(M^2/m^2)
\right]
\nonumber\\
&+&
\frac{\alpha_Y}{4\pi}
m_e \tan\beta
\times
\frac{\partial}{\partial m^2}
\left[
\frac{1}{2 m^2} f_0(|\mu_{\tilde{g}_Y}|^2/m^2)
\right],
\nonumber\\
d_Y^{(e)\prime} (|\mu_{\tilde{g}_Y}|^2)
&=&
\frac{\alpha_Y}{4\pi}
m_e
\times
\frac{\partial}{\partial m^2}
\left[
\frac{1}{2m^2} f_0(|\mu_{\tilde{g}_Y}|^2/m^2)
\right].
\end{eqnarray}
Here, a finite-difference operator and mass functions $f_0$ and $f_1$  are 
defined as
\begin{eqnarray}
\hat{D}^{(x, y)}_{z}
\left[f(z)\right]
&\equiv&
\frac{f(x)-f(y)}{x-y}
\end{eqnarray}
and
\begin{eqnarray}
f_0(x)&=&
\frac{1}{(1-x)^3} (1-x^2+2 x \log x),
\nonumber\\
f_0(x)&=&
\frac{1}{(1-x)^3} (3-4x+x^2 +2  \log x).
\end{eqnarray}

For down quark, the functions in Eq.~(\ref{qedm}) are 
\begin{eqnarray}
d_3^{(d)}(|\mu_{\tilde{g}_3}|^2)
&=&
\frac{\alpha_3}{4\pi}
m_d\tan\beta\times
\frac{\partial}{\partial m^2}
\left[
\frac{4}{9m^2} f_0(|\mu_{\tilde{g}_3}|^2/m^2)
\right],
\nonumber\\
d_2^{(d)}(|\mu_{\tilde{g}_2}|^2)
&=&
\frac{\alpha_2}{4\pi}
m_d\tan\beta\times
\hat{D}^{(|\mu_{\tilde{g}_2}|^2, |\mu_H|^2)}_{M^2}
\left[
\frac{1}{4m^2} f_0(M^2/m^2)
-\frac{1}{2m^2} f_1(M^2/m^2)
\right],
\nonumber\\
d_Y^{(d)}(|\mu_{\tilde{g}_Y}|^2)
&=&
\frac{\alpha_Y}{4\pi}
m_d \tan\beta
\times
\hat{D}^{(|\mu_{\tilde{g}_Y}|^2, |\mu_H|^2)}_{M^2}
\left[
-\frac{1}{12m^2} f_0(M^2/m^2)
\right]
\nonumber\\
&+&
\frac{\alpha_Y}{4\pi}
m_d \tan\beta
\times
\frac{\partial}{\partial m^2}
\left[
-\frac{1}{54m^2} f_0(|\mu_{\tilde{g}_Y}|^2/m^2)
\right],
\nonumber\\
d_3^{(d) \prime}(|\mu_{\tilde{g}_3}|^2)
&=&
\frac{\alpha_3}{4\pi}
m_d\times
\frac{\partial}{\partial m^2}
\left[
\frac{4}{9m^2} f_0(|\mu_{\tilde{g}_3}|^2/m^2)
\right],
\nonumber\\
d_Y^{(d)\prime} (|\mu_{\tilde{g}_Y}|^2)
&=&
\frac{\alpha_Y}{4\pi}
m_d 
\times
\frac{\partial}{\partial m^2}
\left[
-\frac{1}{54m^2} f_0(|\mu_{\tilde{g}_Y}|^2/m^2)
\right],
\end{eqnarray}
and
\begin{eqnarray}
\bar{d}_3^{(d)}(|\mu_{\tilde{g}_3}|^2)
&=&
\frac{\alpha_3}{4\pi}
m_d\tan\beta\times
\frac{\partial}{\partial m^2}
\left[
\frac{1}{6m^2} f_0(|\mu_{\tilde{g}_3}|^2/m^2)
-\frac{3}{2m^2} f_1(|\mu_{\tilde{g}_3}|^2/m^2)
\right],
\nonumber\\
\bar{d}_2^{(d)}(|\mu_{\tilde{g}_2}|^2)
&=&
\frac{\alpha_2}{4\pi}
m_d\tan\beta\times
\hat{D}^{(|\mu_{\tilde{g}_2}|^2, |\mu_H|^2)}_{M^2}
\left[
\frac{3}{4m^2} f_0(M^2/m^2)
\right],
\nonumber\\
\bar{d}_Y^{(d)}(|\mu_{\tilde{g}_Y}|^2)
&=&
\frac{\alpha_Y}{4\pi}
m_d \tan\beta
\times
\hat{D}^{(|\mu_{\tilde{g}_Y}|^2, |\mu_H|^2)}_{M^2}
\left[
\frac{1}{4m^2} f_0(M^2/m^2)
\right]
\nonumber\\
&+&
\frac{\alpha_Y}{4\pi}
m_d \tan\beta
\times
\frac{\partial}{\partial m^2}
\left[
\frac{1}{18m^2} f_0(|\mu_{\tilde{g}_Y}|^2/m^2)
\right],
\nonumber\\
\bar{d}_3^{(d) \prime}(|\mu_{\tilde{g}_3}|^2)
&=&
\frac{\alpha_3}{4\pi}
m_d \times
\frac{\partial}{\partial m^2}
\left[
\frac{1}{6m^2} f_0(|\mu_{\tilde{g}_3}|^2/m^2)
-\frac{3}{2m^2} f_1(|\mu_{\tilde{g}_3}|^2/m^2)
\right],
\nonumber\\
\bar{d}_Y^{(d)\prime} (|\mu_{\tilde{g}_Y}|^2)
&=&
\frac{\alpha_Y}{4\pi}
m_d 
\times
\frac{\partial}{\partial m^2}
\left[
\frac{1}{18m^2} f_0(|\mu_{\tilde{g}_Y}|^2/m^2)
\right].
\end{eqnarray}

For up quark,  
\begin{eqnarray}
d_3^{(u)}(|\mu_{\tilde{g}_3}|^2)
&=&
\frac{\alpha_3}{4\pi}
m_u\tan^{-1}\beta\times
\frac{\partial}{\partial m^2}
\left[
-\frac{8}{9m^2} f_0(|\mu_{\tilde{g}_3}|^2/m^2)
\right],
\nonumber\\
d_2^{(u)}(|\mu_{\tilde{g}_2}|^2)
&=&
\frac{\alpha_2}{4\pi}
m_u\tan^{-1}\beta\times
\hat{D}^{(|\mu_{\tilde{g}_2}|^2, |\mu_H|^2)}_{M^2}
\left[
\frac{1}{2m^2} f_1(M^2/m^2)
\right],
\nonumber\\
d_Y^{(u)}(|\mu_{\tilde{g}_Y}|^2)
&=&
\frac{\alpha_Y}{4\pi}
m_u \tan^{-1}\beta
\times
\hat{D}^{(|\mu_{\tilde{g}_Y}|^2, |\mu_H|^2)}_{M^2}
\left[
\frac{1}{6m^2} f_0(M^2/m^2)
\right]
\nonumber\\
&+&
\frac{\alpha_Y}{4\pi}
m_u \tan^{-1}\beta
\times
\frac{\partial}{\partial m^2}
\left[
-\frac{2}{27m^2} f_0(|\mu_{\tilde{g}_Y}|^2/m^2)
\right],
\nonumber\\
d_3^{(u) \prime}(|\mu_{\tilde{g}_3}|^2)
&=&
\frac{\alpha_3}{4\pi}
m_u\times
\frac{\partial}{\partial m^2}
\left[
-\frac{8}{9m^2} f_0(|\mu_{\tilde{g}_3}|^2/m^2)
\right],
\nonumber\\
d_Y^{(u)\prime} (|\mu_{\tilde{g}_Y}|^2)
&=&
\frac{\alpha_Y}{4\pi}
m_u 
\times
\frac{\partial}{\partial m^2}
\left[
-\frac{2}{27m^2} f_0(|\mu_{\tilde{g}_Y}|^2/m^2)
\right],
\end{eqnarray}
and
\begin{eqnarray}
\bar{d}_3^{(u)}(|\mu_{\tilde{g}_3}|^2)
&=&
\frac{\alpha_3}{4\pi}
m_u\tan^{-1}\beta\times
\frac{\partial}{\partial m^2}
\left[
\frac{1}{6m^2} f_0(|\mu_{\tilde{g}_3}|^2/m^2)
-\frac{3}{2m^2} f_1(|\mu_{\tilde{g}_3}|^2/m^2)
\right],
\nonumber\\
\bar{d}_2^{(u)}(|\mu_{\tilde{g}_2}|^2)
&=&
\frac{\alpha_2}{4\pi}
m_u\tan^{-1}\beta\times
\hat{D}^{(|\mu_{\tilde{g}_2}|^2, |\mu_H|^2)}_{M^2}
\left[
\frac{3}{4m^2} f_0(M^2/m^2)
\right],
\nonumber\\
\bar{d}_Y^{(u)}(|\mu_{\tilde{g}_Y}|^2)
&=&
\frac{\alpha_Y}{4\pi}
m_u \tan^{-1}\beta
\times
\hat{D}^{(|\mu_{\tilde{g}_Y}|^2, |\mu_H|^2)}_{M^2}
\left[
\frac{1}{4m^2} f_0(M^2/m^2)
\right]
\nonumber\\
&+&
\frac{\alpha_Y}{4\pi}
m_u \tan^{-1}\beta
\times
\frac{\partial}{\partial m^2}
\left[
-\frac{1}{9m^2} f_0(|\mu_{\tilde{g}_Y}|^2/m^2)
\right],
\nonumber\\
\bar{d}_3^{(u) \prime}(|\mu_{\tilde{g}_3}|^2)
&=&
\frac{\alpha_3}{4\pi}
m_u \times
\frac{\partial}{\partial m^2}
\left[
\frac{1}{6m^2} f_0(|\mu_{\tilde{g}_3}|^2/m^2)
-\frac{3}{2m^2} f_1(|\mu_{\tilde{g}_3}|^2/m^2)
\right],
\nonumber\\
\bar{d}_Y^{(u)\prime} (|\mu_{\tilde{g}_Y}|^2)
&=&
\frac{\alpha_Y}{4\pi}
m_u 
\times
\frac{\partial}{\partial m^2}
\left[
-\frac{1}{9m^2} f_0(|\mu_{\tilde{g}_Y}|^2/m^2)
\right].
\end{eqnarray}

\newpage
%
%
\newcommand{\Journal}[4]{{\sl #1} {\bf #2} {(#3)} {#4}}
\newcommand{\APJ}{Ap. J.}
\newcommand{\CJP}{Can. J. Phys.}
\newcommand{\MPL}{Mod. Phys. Lett.}
\newcommand{\NC}{Nuovo Cimento}
\newcommand{\NP}{Nucl. Phys.}
\newcommand{\PL}{Phys. Lett.}
\newcommand{\PR}{Phys. Rev.}
\newcommand{\PRep}{Phys. Rep.}
\newcommand{\PRL}{Phys. Rev. Lett.}
\newcommand{\PTP}{Prog. Theor. Phys.}
\newcommand{\SJNP}{Sov. J. Nucl. Phys.}
\newcommand{\ZP}{Z. Phys.}

\end{document}